\documentclass[12pt,english,sort&compress]{elsarticle}
\usepackage[T1]{fontenc}
\usepackage[latin9]{inputenc}
\usepackage{geometry}
\usepackage{color}
\usepackage{array}
\usepackage{units}
\usepackage{amsmath}
\usepackage{amssymb}
\usepackage{graphicx}

\makeatletter

\providecommand{\tabularnewline}{\\}

\usepackage{multirow}
\usepackage {url}
\usepackage[backref]{hyperref}
\hypersetup{colorlinks=false}
\usepackage{graphicx,eso-pic,xcolor}
\makeatletter
\AddToShipoutPicture{%
\setlength{\@tempdimb}{.5\paperwidth}%
\setlength{\@tempdimc}{.5\paperheight}%
\setlength{\unitlength}{1pt}%
}
\makeatother

\makeatother

\usepackage{babel}
\begin{document}

\title{An efficient mixed-precision, hybrid CPU-GPU implementation of a
fully implicit particle-in-cell algorithm}

\author{G. Chen, L. Chacón, }

\address{Oak Ridge National Laboratory, Oak Ridge,TN 37831, USA }

\author{D. C. Barnes}

\address{Coronado Consulting, Lamy, NM 87540, USA }
\begin{abstract}
Recently, a fully implicit, energy- and charge-conserving particle-in-cell
method has been proposed for multi-scale, full-\textit{f} kinetic
simulations {[}G. Chen, \emph{et al.}, \emph{J. Comput. Phys. }\textbf{230},
18 (2011){]}. The method employs a Jacobian-free Newton-Krylov (JFNK)
solver, capable of using very large timesteps without loss of numerical
stability or accuracy. A fundamental feature of the method is the
segregation of particle-orbit computations from the field solver,
while remaining fully self-consistent. This paper describes a very
efficient, mixed-precision hybrid CPU-GPU implementation of the implicit
PIC algorithm exploiting this feature. The JFNK solver is kept on
the CPU in double precision (DP), while the implicit, charge-conserving,
and adaptive particle mover is implemented on a GPU (graphics processing
unit) using CUDA in single-precision (SP). Performance-oriented optimizations
are introduced with the aid of the roofline model. The implicit particle
mover algorithm is shown to achieve up to 400 GOp/s on a Nvidia GeForce
GTX580. This corresponds to $25$\% absolute GPU efficiency against
the peak theoretical performance, and is about 300 times faster than
an equivalent serial CPU (Intel Xeon X5460) execution. For the test
case chosen, the mixed-precision hybrid CPU-GPU solver is shown to
over-perform the DP CPU-only serial version by a factor of $\sim100$,
without apparent loss of robustness or accuracy in a challenging long-timescale
ion acoustic wave simulation.
\end{abstract}
\maketitle

\section{Introduction}

Particle-in-cell (PIC) methods were developed in the 1960's for the
simulation of plasma systems with many particles interacting via electromagnetic
fields. The technique employs the method of characteristics to follow
discrete volumes (finite-size particles) in phase space, with fields
defined on a discrete mesh in physical space. In a typical PIC timestep,
particle orbits are integrated to find new positions and velocities
for given fields. New fields are found by solving Maxwell's equations
(or a subset thereof) using new moments (charge density and/or current)
computed from particles. Interpolation operations are defined to exchange
information between particles and mesh quantities. The method has
been very successful in its application to many areas in plasma physics
\citep{birdsall-langdon} and beyond.

Due to the intrinsic data-parallel nature of the orbit integrals,
PIC methods have been particularly successful in exploiting current
(petascale) supercomputers \citep{liewer1989general,bowers2009advances}.
However, with the current trend toward million-way parallelism, achieving
high performance and high efficiency for PIC simulations on future
supercomputers will be non-trivial. Future parallel computing systems
will place strong constraints on algorithms both in the amount of
memory available and in the cost of accessing it (memory operations
are much slower than processor computations in modern computers, with
the gap most likely enlarging in the near future \citep{millett2011future}).
As a result, memory-bounded algorithms will be more challenged to
utilize the hardware efficiently. Most current PIC time-stepping algorithms
are explicit, with numerical stability constraints limiting work per
particle with a single, small timestep. As a result, explicit PIC
algorithms are typically memory-bounded, and thus are critically affected
by the memory bottleneck.

Nevertheless, there have been fairly successful efforts in porting
explicit PIC algorithms to new computing architectures such as graphics
processing units (GPUs) \citep{decyk2010adaptable,kong2010particle,burau2010picongpu,madduri2009memory,madduri2011gyrokinetic}.
In \citep{decyk2010adaptable}, a 2D electrostatic PIC algorithm is
implemented on a GPU. A particle data structure is proposed to match
the GPU architecture so that the algorithm can be adapted to different
problem configurations. Reference \citep{kong2010particle} describes
a GPU implementation of a 2D fully relativistic, electromagnetic PIC
code, and introduces a strategy to improve performance of a charge-conserving
current deposition scheme (which would otherwise require many conditional
branches). Reference \citep{burau2010picongpu} further describes
an implementation of a 2D fully relativistic, electromagnetic PIC
code on a GPU-cluster via domain decomposition, featuring MPI for
inter-node communication. References \citep{stantchev2008fast,madduri2009memory,madduri2011gyrokinetic}
report efforts to improve the memory efficiency of particle-grid interpolations
and to reduce memory usage in gyrokinetic codes. In Ref. \citep{madduriSC11gyrokinetic},
a speedup of about 2 is reported with a hybrid CPU-GPU simulation
approach vs. a CPU-only one on tens of thousands of nodes.

In this study, we focus on implementing a novel fully implicit PIC
algorithm for electrostatic simulation\citep{Chen20117018} on a heterogeneous
CPU-GPU architecture. Unlike explicit PIC, the fully implicit PIC
algorithm does not feature a numerical stability timestep constraint.
The method can solve either the coupled Vlasov-Ampère (VA) or Vlasov-Poisson
(VP) system of equations, discretized to feature exact local charge
and global energy conservation theorems. A Jacobian-free Newton-Krylov
(JFNK) nonlinear solver is used to converge the nonlinear residual
to a tight nonlinear tolerance. Crucial to the implicit PIC algorithm
is the concept of particle enslavement, whereby the integration of
particle orbits is an auxiliary computation, segregated in the evaluation
of the nonlinear residual. Particle enslavement has several advantages.
Firstly, it results in much smaller residual vectors (and therefore
in a much reduced memory footprint) because only fields are dependent
variables in the solver. Secondly, it affords one significant freedom
to perform the particle orbit integration. In Ref. \citep{Chen20117018},
this freedom was exploited to implement a self-adaptive, sub-stepping,
charge-conserving particle mover.

Despite the advantages of particle enslavement, the particle integration
step remains the most expensive element in the fully implicit PIC
algorithm. This is so because particle orbits need to be computed
for every nonlinear residual evaluation, and many such residuals are
computed as JFNK converges to a solution. This issue is exacerbated
further by the ability of implicit PIC to use large timesteps, which
requires many orbit integration sub-steps per timestep in the particle
mover. 

Being the most time-consuming operation in the fully implicit PIC
algorithm, the particle orbit integration is the obvious target for
hardware acceleration. Our implementation exploits the flexibility
afforded by particle enslavement, and targets the particle orbit integration
step for the GPU, while the field solver remains on the CPU. Contrary
to most PIC algorithms, we show that the implicit particle mover is
compute-bounded, and therefore has the potential of efficiently utilizing
both the extreme multi-threading capability and the large operational
throughput of the GPU architecture. Communication between CPU and
GPU routines involves particle moments only (which are grid quantities),
and not the particles themselves, thus minimizing the impact of memory
bandwidth bottlenecks. 

We use the roofline analytical performance model\citep{williams2009roofline}
to help understand the performance limitations and bottlenecks of
the algorithm and achieve high performance on the GPU. There are several
design constraints for optimizing code on a GPU \citep{ryoo2008optimization,hwu2009compute,kirk2010programming,nvprogramming,nvbestpracticesguide}:
\begin{itemize}
\item Global memory operations are very expensive, and can severely limit
the throughput of the simulation.
\item Not all arithmetic operations are equally fast. The slow operations,
such as square root and division, can hinder the computational throughput.
\item CUDA employs a lockstep execution paradigm within a warp, which comprises
32 threads. Divergent control flow is allowed, but results in performance
degradation.
\item Memory collisions occur when many threads in parallel try to access
the same memory location at the same time. Resolving them serializes
the code and can become the bottleneck in parallel computations. 
\end{itemize}
To mitigate the impact of these constraints on GPU performance, we
have implemented a series of thread-level and warp-level optimizations
in the particle orbit computation without sacrificing accuracy. As
a result of these optimizations, our implicit particle mover achieves
up to 300-400 GOp/s for VA and VP, respectively, on an Nvidia GeForce
GTX580, with the VP approach performing better on account of the memory-collision
issue. The corresponding GPU efficiency is $20-25$\% of peak performance.
The accuracy and performance of the overall hybrid CPU-GPU implicit
PIC algorithm is demonstrated using a challenging, long-timescale
ion acoustic wave (IAW) simulation. It is shown that a mixed-precision
implementation, in which the CPU JFNK code uses double-precision and
the GPU particle mover code uses single-precision, can be sufficient
for accuracy. For the test case chosen, this setup results in speedups
of the hybrid CPU-GPU algorithm vs. the CPU-only one up to $100$.
A defect-correction approach has also been implemented that enables
the hybrid algorithm to deliver double-precision results. In this
case, about a third of the GPU calls per time step are made in double
precision, and the speedup drops to $\sim40$. This should be compared
to a factor of 25 speedup obtained when all GPU calls are made in
double precision. These speedups are consistent with Amdahl's law,
as the particle computation takes $\gtrsim98\%$ of the overall computation
time for the test case chosen. 

The rest of the paper is organized as follows. Section \ref{sec:gpu performance model}
introduces the Nvidia GPU Fermi architecture and the roofline model.
Section \ref{sec:ACC-CN mover} describes the specific GPU optimizations
introduced in the adaptive, charge-conserving particle mover. Section
\ref{sec:Num-Examples} shows the performance and efficiency results
of numerical experiments, including the complete IAW test case integrated
with the JFNK solver. Finally, we provide some discussion and conclusions
in Sec. \ref{sec:Conclusion}.

\section{An analytical performance model for GPU computing}

\label{sec:gpu performance model}

As modern computer architectures shift from single- to multi-core
or many-core processors, parallel programs must be able to exploit
increasingly large concurrency efficiently. This, in turn, places
strong emphasis on identifying performance bottlenecks and sources
of latencies. This task is facilitated when programmers have some
basic understanding of the underlying hardware, such as the memory
hierarchy and the processor computing capabilities, and target the
optimizations for that hardware. In this study, we focus on the GPU
architecture, which we introduce next.

\subsection{Nvidia GPU architecture }

Contemporary Nvidia GPUs\citep{nickolls2010gpu} are capable of performing
scientific computations programmed in high-level languages such as
CUDA C/C++ and CUDA Fortran, with high accuracy (supporting IEEE standards)
and high performance (theoretical throughput over trillion floating-point
operations per second or TFLOPS). GPUs consist of many processing
units. For instance, the newest Nvidia GPU to date, named Fermi, has
up to 16 streaming multi-processors (SM). Each SM contains 32 processors,
or CUDA cores, which perform floating-point, integer, and logic operations.
SMs also contain 4 special function units (SFU), which calculate fast
floating-point approximations to certain complex operations such as
reciprocal, reciprocal square root, etc.

GPUs also contain its own memory system. From slow to fast, one finds
off-chip global memory, on-chip shared memory, and on-chip register
files. In addition, Fermi GPUs are equipped with a two-level (L1 and
L2) read/write cache hierarchy. Specifically, Fermi GPUs contains
2 to 6 GB global memory and 768 KB L2 cache; each SM contains 64 KB
configurable shared memory/L1 cache, and 128 KB registers. A unique
feature of the GPU memory system is that all levels (except for the
L2 cache) can be explicitly managed by the programmer.

In order to gain insight into the maximum performance and efficiency
of a given algorithm running on a GPU, we adopt an analytical performance
model, the so called roofline model\citep{williams2009roofline}.
We proceed to introduce the roofline model and its application for
the Nvidia GPU architecture. In the sequel, we assume that all computational
operations are on 32-bit words (e.g., single-precision) unless otherwise
specified.

\subsection{Roofline model\label{sub:Roofline-model}}

The roofline model is motivated by the fact that the bandwidth of
current computer off-chip memory is often much slower than the throughput
of the processing unit\citep{wulf1995hitting}. For instance, the
Nvidia GeForce GTX580 GPU has a DRAM bandwidth of 192 GB/s, whereas
its peak floating-point operational throughput is 1581 GFLOPS. This
large discrepancy makes identifying whether the program is memory-bounded
or compute-bounded critical to target optimizations. If memory-bounded,
the program should maximize the use of fast memory; if compute-bounded,
it should minimize the number of operations.

The roofline model provides a simple method to determine whether an
algorithm is either memory-bounded or compute-bounded. The key figure
of merit is the operational intensity (OI), defined as the ratio of
compute operations to memory operations. An algorithm is compute-bounded
for OI higher than the balanced OI, and is memory-bounded otherwise.
The balanced OI of the target device is defined as the ratio of peak-operational
throughput to memory bandwidth (which is about 8 FLOP/B for the Nvidia
GeForce GTX580 GPU). 

An algorithm's compute efficiency is commonly defined as the ratio
of its FLOPS vs. the peak theoretical compute performance. On a Nvidia
GPU, the latter is calculated as 
\begin{equation}
(\mathrm{number\, of}\,\mathrm{cores})\times2\,\mathrm{Flop/CC}\times(\mathrm{clock\, rate)},\label{eq:theoreticalPeak}
\end{equation}
where the factor of 2 comes from the fused multiply-add (FMA) operation,
which computes one AXPY operation (i.e., $a\times x+y$) per clock
cycle (CC). On the GeForce GTX580 GPU, $\mathrm{number\, of\, cores}=512$,
and $\mathrm{clock\, rate=1.544}$ GHz, resulting in 1.58 TFLOPS.
Note that, by definition, the peak theoretical performance may be
reached by algorithms based on FMA operations only. However, most
algorithms in scientific computing mix floating-point, integer, and
logic operations. Such algorithms \textit{cannot} reach the peak theoretical
performance, no matter how well optimized. Therefore, for a given
algorithm, it is useful to define an intrinsic efficiency based on
its specific operations. 

\textcolor{black}{We define the }\textcolor{black}{\emph{theoretical
operational throughput}}\textcolor{black}{{} as the maximum theoretical
performance of a compute-bounded algorithm. It can be calculated as
\begin{equation}
(\mathrm{number\, of}\,\mathrm{SM})\times(\mathrm{average\, operational\, throughput)}\times(\mathrm{clock\, rate)},\label{eq:ceiling-line}
\end{equation}
where 
\begin{equation}
(\mathrm{average\, operational\, throughput)}\equiv\sum(\mathrm{operations)}/\sum(\mathrm{clock\, cycles})\label{eq:avg_op_thput-1}
\end{equation}
is the average number of operations per clock cycle per stream multi-processor.
The theoretical operational throughput assumes that all memory and
instruction latencies are completely hidden or negligible, without
performance overhead of any kind. The }\textcolor{black}{\emph{intrinsic
efficiency is}}\textcolor{black}{{} defined as the ratio of the actual
operational throughput vs. the theoretical one} (Eq. \ref{eq:ceiling-line}).
It indicates an algorithm's real effectiveness in using a given target
hardware.

In what follows, \textcolor{black}{we compare the operational throughput
of several basic operations, which are the building blocks of our
particle mover algorithm, on the Nvidia Geforce GTX580 GPU in the
context of the roofline model. }

\begin{figure}
\includegraphics[width=\columnwidth]{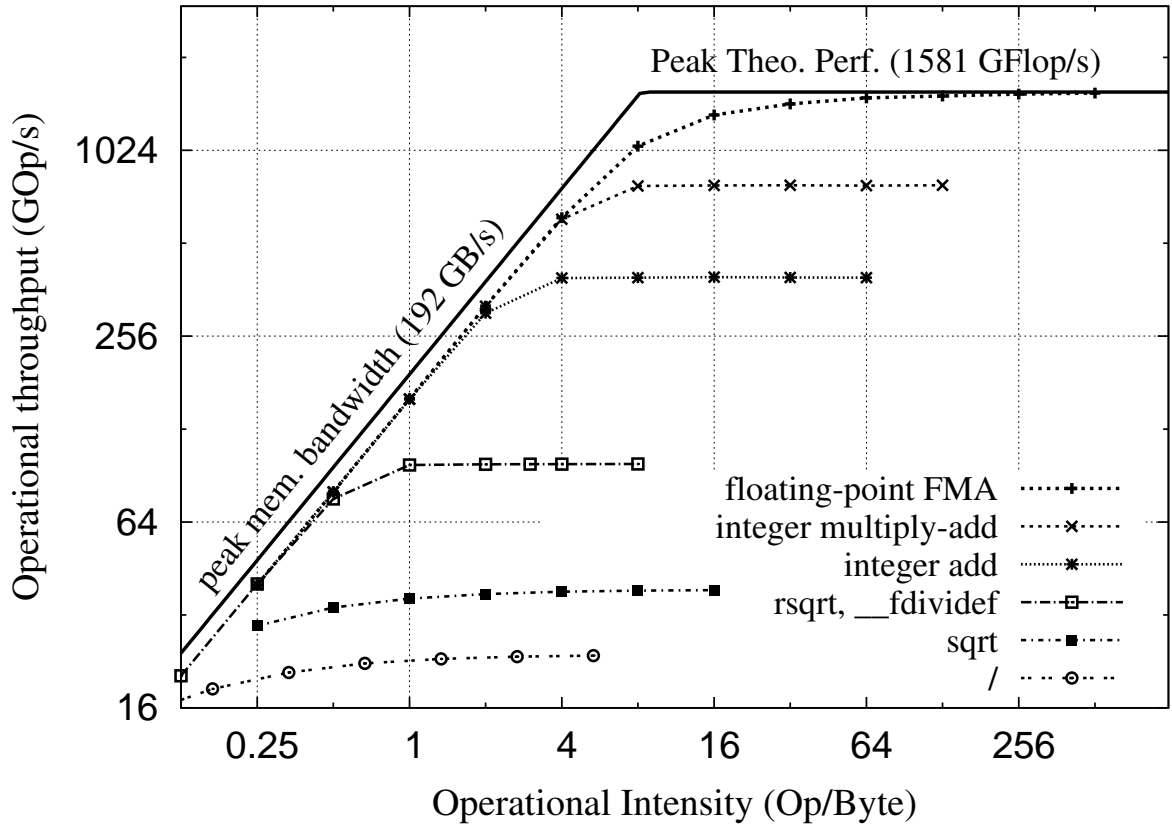}\caption{\label{fig:Roofline-ops}Roofline model for Nvidia GeForce GTX580
on a $\mathrm{log}_{2}$-$\mathrm{log}_{2}$ scale. The performance
lines are obtained by artificial algorithms which repeat a single
operation multiple times. Saturation of the curves shows the maximum
operational throughput in the compute-bounded regime for a given operation.
In the plot, sqrt, /, rsqrt, and \_\_fdividef stand for IEEE compliant
square root, IEEE compliant division, intrinsic reciprocal square
root, and intrinsic fast division, respectively.}
\end{figure}

\subsection{Operational throughput \label{sub:Operation-throughput}}

Each CUDA SM of the Nvidia GPU Fermi GF100 architecture\citep{wittenbrink2011fermi}
has 32 floating-point units (FPUs), 32 integer arithmetic logic units
(ALUs), and 4 SFUs, capable of processing different types of computations
simultaneously. Depending on the hardware implementations, the throughput
of different computational operations varies\citep{nvprogramming}.
To illustrate this, we have created a simple CUDA code to micro-benchmark
(similar to those used in Ref.\citep{wong2010demystifying,volkov2008benchmarking})
the throughput of some basic floating-point and integer operations.
In these tests, many identical operations (using unrolled loops) are
performed by each thread, and many concurrent threads (with 100\%
occupancy) are employed. The test code is compiled with nvcc v4.0
compiler. The assembly code generated by the PTX (or cuobjdump) tool
is examined to ensure that instructions are executed as intended.
Results are shown in Fig.\ref{fig:Roofline-ops}, from which we can
make the following observations:
\begin{itemize}
\item As expected, FMA reaches the peak theoretical performance for large
OIs.
\item The peak performance of integer operations (multiply-add) is at best
half of the peak theoretical performance. In other words, integer
operations are slower than floating-point operations. 
\item Standard (IEEE compliant) square root (sqrt) and division(/) operations
are one to two orders of magnitude slower than simpler floating-point
or integer operations. 
\item The throughput of intrinsic reciprocal square root (rsqrt), or intrinsic
fast division in single-precision (\_\_fdividef), hardware-implemented
in SFUs, is exactly 16 times lower than the peak theoretical performance.
This is expected, as it equals the ratio of the number of FPU (32)
and SFU(4) multiplied by 2 (operations per FMA). Intrinsic operations
are much faster than IEEE versions at the cost of being less accurate.
We will show in later sections that, under some situations, fast intrinsic
functions can be accompanied by a few FMA operations to improve accuracy.
The hybrid use of SFUs and FPUs has the advantage of exploiting both
units concurrently, and therefore offers the potential to achieve
sufficient accuracy with enhanced performance.
\end{itemize}
Table \ref{tab:Throughput-GTX580} lists the maximum throughput of
the operations on the GPU found by the micro-benchmarks. With the
throughput of each basic operation known, we can now calculate the
average operational throughput for a particular algorithm {[}\textcolor{black}{Eq.\eqref{eq:avg_op_thput-1}{]},
with }the total number of clock cycles required to execute those operations
calculated as $\sum_{i}N_{i}\times32/OT_{i}$. Here, $N_{i}$ is the
number of operations of type $i$, and $OT_{i}$ is the operational
throughput from Table \ref{tab:Throughput-GTX580}\textcolor{black}{. }

\textcolor{black}{We proceed to analyze in detail the GPU implementation
of the implicit particle mover of Ref.\citep{Chen20117018}}.

\begin{table}
\begin{centering}
\caption{\label{tab:Throughput-GTX580}Throughput of floating point (FP), arithmetic
and logic (AL), and special function (SF) operations on GeForce GTX
580. Throughput is measured as operations per clock cycle per multiprocessor.
The compiler nvcc v4.0 is used to produce the results.}

\par\end{centering}

\centering{}%
\begin{tabular}{|>{\centering}m{1in}|>{\centering}m{1.5in}|>{\centering}m{0.8in}|>{\centering}m{0.4in}|}
\hline 
\noalign{\vskip\doublerulesep}
FP add,mul, (fma) & AL add, mul, logic, cvt (mul-add) & SF rsqrt, \_\_fdividef & $/$\tabularnewline[\doublerulesep]
\hline 
32(64) & 16(32) & 4 & $\lesssim$1\tabularnewline
\hline 
\end{tabular}
\end{table}

\section{GPU implementation and performance analysis of the adaptive, charge-conserving
particle mover}

\label{sec:ACC-CN mover}

We pursue the development of a hybrid CPU-GPU algorithm in which the
particle push is farmed off to the GPU, while the nonlinear solver
remains on the CPU. In our implementation, particles reside in the
GPU global memory, and no particle transfer is needed between CPU
and GPU for the field solver. Only grid quantities (i.e., the electric
field and moments collected from the particles) are transferred between
the two architectures. Nonlinear iterations are performed on the CPU
until nonlinear convergence is achieved (convergence is enforced on
field quantities, with particle quantities obtained self-consistently).
The process is repeated every timestep.

This section focuses on the implicit particle mover \citep{Chen20117018},
which features three main properties: self-adaptivity, automatic local
charge conservation, and absolute numerical stability. Self-adaptivity
is achieved by particle sub-stepping, with the sub-timestep controlling
the discretization errors of the orbit integral. In each sub-step,
particles are pushed along a straight line. Local charge conservation
is automatically enforced by ensuring particles land at cell boundaries
during the orbit integration process. A Crank-Nicolson (CN) integration
scheme, which is time-centered and implicit, is employed to guarantee
numerical stability for arbitrary timesteps. 

It is well known that contemporary GPUs are capable of launching a
large number of threads (tens of thousands on current-generation devices)
in a SPMD (single program, multiple data) style. They are very attractive
for particle simulations, as particle orbits are independent of each
other. However, the highly dynamic nature of the implicit mover may
prevent the algorithm from achieving high performance and efficiency
on the GPU. To begin with, the simulated systems are typically highly
nonlinear and inhomogeneous. As a result, threads pushing particles
with different positions in phase space will experience very different
workloads. When large timesteps are employed, each particle follows
an orbit according to local conditions, undergoing an indefinite number
of sub-timesteps and cell-crossings. Similarly, an iterative solution
of the coupled Crank-Nicolson equations introduces unpredictability
in the algorithm, for the number of iterations for convergence is
unknown and particle-dependent. The resulting unpredictable logical
paths (per thread) create divergent branches, serializing the executions
in a given warp (32 threads). Non-divergent branches may also be created,
with some idle threads waiting for others to finish. It follows that
it may be very difficult to keep all threads busy, resulting in parallel
performance degradation. 

Additional performance degradation can result from inefficient memory
operations. One prime example is parallel scatter-gather operations,
such as field interpolations to particles and moment integration from
particles, which can significantly slow down the simulation. Random
access to the field may have a large performance penalty, for instance,
if the L1 cache miss-rate is high. Moment accumulation may hinder
the computation due to the cost of resolving memory collisions. This
occurs when two or more threads try to write the same memory location
simultaneously. Ensuring correctness requires using atomic operations,
which serialize the accumulations.

Despite these challenges, we demonstrate in the following sections
that the implicit particle mover algorithm can in fact be relatively
efficient on the GPU. With the aid of the roofline model, we identify
that the implicit particle mover is compute-bounded, and therefore
significant performance improvement can be achieved through targeted
optimizations. We motivate specific optimizations with a baseline
GPU implementation of the mover (described below), aiming at minimizing
clock cycles (per thread) without accuracy degradation. We also describe
warp-level optimizations, including our particle-sorting strategy
and the use of a warp vote function, to minimize load imbalance and
control-flow latencies. 

Before getting into the detailed optimizations of the mover, we introduce
its basic memory management. There are mainly two groups of memory
operations. The first group involves reading particle quantities \{$v_{p}$,$x_{p}$,$i_{p}$\}
(denoting velocity, position, and cell index, respectively) at the
timestep $n$ from global GPU memory, and writing the updated quantities
at the timestep ($n+1$) to global GPU memory. Since the particle-orbit
calculations are independent of each other, the load and store of
the particle quantities, which we group in a structure-of-array fashion,
are trivially coalesced (making stride-one access of the global memory
for optimal performance\citep{nvbestpracticesguide}). The second
group involves reading/writing grid quantities, such as the electric
field $E$, current density $j$, or charge density $\rho$. The $E$
field is read-only; either L1 cache or texture memory can be exploited
to accelerate repeated readings. Our approach to the accumulation
of moments (such as $j$ or $\rho$), which is often found to be the
bottleneck of many PIC implementations \citep{stantchev2008fast,madduri2009memory,burau2010picongpu},
is discussed in detail below (Sec. \ref{sub:Particle-current-deposition}). 

To frame the subsequent discussion, we divide each particle sub-step
into four algorithmic elements:
\begin{enumerate}
\item Estimate the sub-timestep $\Delta\tau$. 
\item Integrate the orbit over $\Delta\tau$ using a Crank-Nicolson scheme.
\item If a particle crosses a cell boundary, make it land at the boundary.
\item Accumulate the current density on the grid points (for VA, but not
for VP).
\end{enumerate}
This process is repeated over many sub-steps until the end of the
timestep, at which point we either take the orbit average of the current
density (for VA), or accumulate the charge density (for VP). In the
following, we describe the baseline algorithm and targeted optimizations
for each element.

\subsection{Algorithmic element $\#$1: estimate of the sub-timestep\label{sub:Algorithmic-element-1-sub-time}}

The first algorithmic element employs a standard local-error-control
method\citep{shampine2005error} to estimate the sub-timestep, by
taking the difference between the truncation errors of the Euler's
scheme (a first-order method) and Heun's scheme (a second-order method)
to be smaller than a specified tolerance. The resulting formula reads\citep{Chen20117018}

\begin{equation}
\left\Vert le(\Delta\tau)\right\Vert _{2}<\varepsilon_{a}+\varepsilon_{r}\left\Vert r^{0}(\Delta\tau)\right\Vert _{2},\label{eq:localerror}
\end{equation}
where $\left\Vert \cdot\right\Vert _{2}$ denotes the $L_{2}$-norm
of enclosed vector, $le(\Delta\tau)=\frac{(\Delta\tau)^{2}}{2}\{a_{p}^{\nu},\left(\frac{\partial a_{p}}{\partial x}v_{p}\right)^{\nu}\}$
is the local truncation error of the sub-timestep $\nu$, $\varepsilon_{a}$
and $\varepsilon_{r}$ are absolute and relative tolerances, respectively,
and $r^{0}(\Delta\tau)\equiv\{v_{p}^{\nu},a_{p}^{\nu}\}\Delta\tau$
is the initial residual. For $\partial a_{p}/\partial x$, we take
\begin{equation}
\partial a_{p}/\partial x\cong\frac{q}{m}(E_{i}-E_{i-1})/\Delta x,\label{eq:dadx-approx}
\end{equation}
when the particle is in cell $i$ (or between grid points $i-1$ and
$i$) at time level $\nu$. This is exact for linear interpolations.
It follows that the estimate can be found by solving a quadratic equation
for $\Delta\tau$.

A direct implementation of the above method would require about 33
floating-point operations, 6 special function operations, and 1 division
operation. To optimize the method, we replace the $L_{2}$-norm with
the $L_{1}$-norm, e.g., $\left\Vert \mathbf{\mathit{\mathrm{\{}x_{1},x_{2}\mathrm{\}}}}\right\Vert _{1}=|x_{1}|+|x_{2}|$,
which suffices for the error estimate without requiring a square root.
The equation for $\Delta\tau$ {[}from Eq.\eqref{eq:localerror}{]}
can be written as
\begin{equation}
\alpha\Delta\tau^{2}-\beta\Delta\tau-\gamma^{2}=0,\label{eq:dtau-quadEq}
\end{equation}
where $\alpha=\left(\left|a_{p}^{\nu}\right|+\left|\left(\frac{\partial a_{p}}{\partial x}v_{p}\right)^{\nu}\right|\right)$/2,
$\beta=\varepsilon_{r}\left(\left|a_{p}^{\nu}\right|+\left|v_{p}^{\nu}\right|\right)$,
$\gamma^{2}=\varepsilon_{a}$. To optimize the computation further,
we avoid solving the quadratic equation by noting that it is sufficient
to estimate $\Delta\tau$ from the absolute and relative tolerances
separately:
\begin{eqnarray}
\alpha\Delta\tau^{2} & = & \gamma^{2},\label{eq:dtau-absle}\\
\alpha\Delta\tau & = & \beta,\label{eq:dtau-relle}
\end{eqnarray}
and then take $\Delta\tau=\mathrm{max}(\gamma d,\beta d^{2})$, with
$d=\alpha^{-\nicefrac{1}{2}}$ computed via the intrinsic rsqrt function
only once.

As a result of these optimizations, the estimate of sub-timestep requires
30 floating-point operations (including 6 FMAs) and 1 special-function
operation. The optimized algorithm has reduced the per-thread clock
cycles from 106 to 28, largely achieved by replacing three square
roots and one division with just one reciprocal square-root operation.

\subsection{Algorithmic element $\#$2: Crank-Nicolson particle move\label{sub:Algorithmic-element-2-CN}}

The second element of the mover is a CN step using the estimated sub-timestep
$\Delta\tau$:
\begin{eqnarray}
\frac{x_{p}^{\nu+1}-x_{p}^{\nu}}{\Delta\tau^{\nu}} & = & v_{p}^{\nu+\nicefrac{1}{2}},\label{eq:xp-nu}\\
\frac{v_{p}^{\nu+1}-v_{p}^{\nu}}{\Delta\tau^{\nu}} & = & a_{p}^{\nu+\nicefrac{1}{2}},\label{eq:vp-nu}
\end{eqnarray}
where $\nicefrac{1}{2}$ denotes a mid-point average, i.e., $v_{p}^{\nu+\nicefrac{1}{2}}\equiv\left(v_{p}^{\nu}+v_{p}^{\nu}\right)/2$.
These equations are implicit and coupled (but not stiff), and are
generally solved by a fixed-point iterative method, e.g., Picard's
method. 

It turns out that, when employing first-order B-spline interpolations,
Eq.\eqref{eq:xp-nu} and \eqref{eq:vp-nu} can be solved directly
as (omitting the subscript \textit{p}) 
\begin{equation}
v^{\nu+1}=\frac{a^{\nu}\Delta\tau^{\nu}+\left[1+\left(\frac{\partial a}{\partial x}\frac{(\Delta\tau)^{2}}{4}\right)^{\nu}\right]v^{\nu}}{1-\left(\frac{\partial a}{\partial x}\frac{(\Delta\tau)^{2}}{4}\right)^{\nu}},\label{eq:vnp-direct}
\end{equation}
where $\partial a/\partial x$ is given by Eq.\eqref{eq:dadx-approx}.
The division in Eq.\eqref{eq:vnp-direct} can be replaced by the intrinsic
reciprocal function without loss of accuracy, as follows. We write
$v^{\nu+1}=NR=N\times\mathrm{RCP}(D)$ where N, D stand for the numerator
and denominator in Eq.\eqref{eq:vnp-direct}, respectively, and $\mathrm{\mathit{R}\equiv RCP}(D)$
stands for the fast intrinsic reciprocal function of CUDA. Note that
the maximum ULP (unit in the last place) error of the fast reciprocal
is about 1\citep{hennessy1994computer} (only slightly lower than
IEEE required 0.5 ULP precision). This is precise to the 8th significant
digit, as the last bit of the single precision number corresponds
to $5.96\cdot10^{-8}$. To reduce the error further, we take one more
iteration of Newton-Raphson's method such that $v^{\nu+1}=NR(2-DR)$\citep{markstein2000ia}. 

The optimization introduced here involves two steps: it first eliminates
the control-flow condition needed in the convergence test of the nonlinear
iteration, and then replaces the division by the intrinsic fast reciprocal
function. In the latter step, one extra step of Newton-Raphson is
taken to ensure sufficient precision. The optimization slightly increases
the number of operations from 11 to 14, but the number of clock cycles
of this step is significantly reduced from 43 to 22.

\subsection{Algorithmic element $\#$3: Particle cell-crossing }

The third algorithmic element checks whether the particle has moved
into another cell after the CN move. If a crossing occurs, the particle
is forced to stop at the cell boundary. The corresponding sub-timestep
is found by solving the quadratic equation:
\[
F(\Delta\tau)=\frac{a^{\nu+\nicefrac{1}{2}}}{2}\Delta\tau^{2}+v^{\nu}\Delta\tau-\Delta x^{\nu}=0,
\]
which is obtained by fixing the final particle position at the boundary
in question, and combining Eq. \eqref{eq:xp-nu} and \eqref{eq:vp-nu}.
The corresponding particle velocity can be found according to the
energy principle to be: 
\begin{equation}
v^{\nu+1}=\mathrm{sng}(\Delta x^{\nu})\sqrt{\left(v^{\nu}\right)^{2}+2a^{\nu+\nicefrac{1}{2}}\Delta x^{\nu}},\label{eq:vnp-sqrt}
\end{equation}
where $\mathrm{sng(\Delta\mathit{x}_{\mathit{}}^{\nu})}$ returns
the sign of $\Delta x^{\nu}(=x^{\nu+1}-x^{\nu})$, which signals the
direction of particle motion. 

The above treatment requires a square root and a division, which can
be optimized by the following (inexact) Newton's method. The solution
is first approximated by 
\begin{equation}
\Delta\tau_{0}=\frac{-v^{\nu}+\mathrm{sng}(\Delta x^{\nu})\sqrt{\left(v_{p}^{\nu}\right)^{2}+2a_{p}^{\nu+\nicefrac{1}{2}}\Delta x_{p}^{\nu}}}{a^{\nu+\nicefrac{1}{2}}}\label{eq:dtau-cc}
\end{equation}
using fast intrinsic functions rsqrt and \_\_fdividef. Subsequent
iterations are performed as:
\begin{equation}
\Delta\tau_{k}=\Delta\tau_{k-1}-F(\Delta\tau_{k-1})/F^{\prime}(\Delta\tau_{k-1}),\label{eq:dtau-Newton_step}
\end{equation}
where $F^{\prime}(\Delta\tau_{k-1})=a^{\nu+\nicefrac{1}{2}}\Delta\tau_{k-1}+v^{\nu}$
is the Jacobian. A fast division can be applied to update Eq. \eqref{eq:dtau-Newton_step}.
This is a safe approximation because the convergence of Newton's method
is robust against small errors in the Jacobian\citep{buttari2007exploiting}.
Note that each of the applied intrinsic functions has a maximum error
of 2 ULP\citep{nvprogramming}, which provides excellent initial value
for Newton iterations. Because of the quadratic convergence rate of
Newton's method (i.e., the correct digits double for every iteration),
this last step ensures that the solution is accurate to the last digit
of a single-precision number.

Overall, we have replaced a square root and a division by a fast reciprocal
square root and two fast divisions. Consequently, the number of clock
cycles for particles moving inside the cell is reduced from 74 to
47.

\subsection{Algorithmic element $\#$4: particle current/charge accumulation\label{sub:Particle-current-deposition}}

In the VA approach, the fourth element employs a standard interpolation
procedure to accumulate the current density on the grid points from
particles:
\[
j=\frac{1}{\Delta x}\sum_{p}q_{p}v_{p}^{\nu+\nicefrac{1}{2}}S(x-x_{p}^{\nu+\nicefrac{1}{2}}),
\]
where $\Delta x$ is the cell size, and $S$ is the shape function.
This is done for every sub-timestep. In the VP approach, the charge
density is collected from each particle as:

\[
\rho=\frac{1}{\Delta x}\sum_{p}q_{p}S(x-x_{p}^{n+1}).
\]
This is done only at the end of the orbit computation. In the baseline
implementation, all the accumulations are performed on shared memory,
using the floating-point atomicAdd function, and the final results
are written back to global memory at the end of the timestep.

It is desirable to avoid memory collisions as much as possible in
the accumulation process. On one hand, memory collisions are completely
avoided when each thread accesses its own copy of the physical domain
in memory, but this is very demanding memory-size-wise. On the other
hand, memory-size requirements are minimized when all threads (of
one thread block) access the same memory domain, but this results
in frequent memory collisions. To balance best efficiency of the accumulation
and limited resources of shared memory, we provide each warp with
its own memory domain\citep{shams2007efficient}. This avoids memory
collisions between threads of different warps. Collisions within a
warp are resolved by the shared-memory, floating-point atomicAdd function. 

A second optimization (for VA only) replaces the shared-memory accumulations
of local current density in a given cell by register accumulations.
This optimization exploits the fact that register access is faster
than shared memory\citep{nvbestpracticesguide}, and collisions with
register accumulations are absent. Specifically, each thread employs
two local register variables (one per cell face) to accumulate current
density as the particle sub-steps within a cell. Register values are
atomically added to shared memory and reset to zero only when the
particle crosses a cell boundary.

Acceleration has been achieved in the moment accumulation by reducing
memory collisions (for both VA and VP), and by reducing the usage
of atomic operations (for VA). Quantitative results are presented
in the following sections.

\subsection{Roofline analysis \label{sub:Roofline-analysis-baseline}}

We proceed to show that the implicit particle mover algorithm is compute-bounded.
The analysis is carried out for VA for brevity. A similar analysis,
with similar conclusions, applies to VP. 

The particle mover algorithm requires two (read/write) single-precision
memory transfer operations from global memory of three particle quantities
\{$x,v,i$\} per particle orbit integration (recall that the electric
field is read-only, and is cached for fast access). The corresponding
(global) memory throughput is $2\times3\times4=24$ B per particle.
For the sake of argument, we assume 20 sub-steps and 2 cell-crossings
in a typical particle orbit per timestep, which results in 1762 (1410)
computational operations per particle in the baseline (optimized)
algorithm. It follows that the corresponding operational intensity
is: 
\[
\mathrm{OI_{im}}\simeq\begin{cases}
73 & \mathrm{[baseline]}\\
59 & [\mathrm{optimized}]
\end{cases}\mathrm{(Op/B)}.
\]
This is 7 to 9 times larger than the balanced OI (see Sec. \ref{sub:Roofline-model}),which
confirms that the implicit particle mover algorithm is compute-bounded.

\subsection{Particle sorting strategy}

It is often beneficial to sort particles, as the memory operations
are more efficient with improved data reuse. Previous studies\citep{bowers2001accelerating,stantchev2008fast,kong2010particle,decyk2010adaptable}
have focused on explicit PIC algorithms. Explicit schemes employ small
timesteps, so that only a small fraction of the particles that cross
cells (or sub-domains) need sorting. The goal is to keep the memory
space consistent with the physical space such that particles that
are physically close are also co-located in memory. Those particles
moving across cells (or sub-domains) need to be rearranged in the
particle array to preserve memory collocation. The complexity of this
process is $O(\eta N)$\citep{kong2010particle}, where $N$ is the
total number of particles and $\eta$ is fraction of the particles
crossing cell (or sub-domain) boundaries. This approach is appropriate
for explicit schemes, where the timestep is very small.

Explicit PIC sorting strategies, however, are not suitable for implicit
PIC when a large timestep is used. In this case, many particles may
cross cells in an implicit timestep, thus increasing the overhead
of shuffling particles in memory and resulting in frequent memory
collisions. In addition, particles with different velocities require
different amount of work per orbit integration, which causes load
imbalances and divergent branches. Hence, in an implicit PIC context,
particle sorting should focus on improving thread load-balancing and
on minimizing memory collisions. We address the former by sorting
particles such that each warp deals with particles with similar velocities
(in sign and magnitude). The latter is addressed by having each thread
within a warp integrate a particle located in a separate cell in physical
space. 

Our particle sorting approach features a two-pass implementation.
In a first pass, we divide the 2D phase space ($x-v$) into rectangular
cells using a Cartesian grid. We label each cell with an integer number
(starting from zero) via lexicographic ordering along the physical
coordinate (rows). Assuming that the physical domain is discretized
with $N_{g}$ cells, the first row in phase space (corresponding to
the same velocity interval) will be labeled with numbers 0,1,...,$N_{g}-1$.
The second row (corresponding to the next velocity interval) will
be labeled with numbers $N_{g}$,...,$2N_{g}-1$, and so on. Next,
we label particles in each cell with the corresponding cell number.
We then collect the total number of particles within each cell, and
use a prefix sum \citep{harris2007parallel} to aggregate particles
in previous cells lexicographically. This gives:
\begin{equation}
N_{i0}=\sum_{j=0}^{i-1}N_{j},
\end{equation}
where $N_{j}$ is the number of particles in cell $j$, and $N_{i0}$
is the total number of particles up to (but excluding) cell $i$.
Here, $i,j$ are lexicographic indexes, and $N_{00}=0$. In a second
pass, we sort particles according to their velocities, and we place
the particles into a 1D array in which each continuous and aligned
$N_{g}$ particles belong to consecutive $N_{g}$ cells in physical
space ($x$), and also have similar velocities. The key in this second
pass is to reserve the particle locations in the 1D array from information
collected during the first pass. In particular, $N_{i0}$ provides
the memory address (starting from index zero) of the 1D particle array
for the first particle in cell $i$. Additional particles in the same
cell with particle index $p\in[1,N_{j}-1]$ are placed in the 1D array
with memory address $N_{i0}+N_{g}\times p$.

When particles are sorted in this way, memory collisions within a
warp are avoided as long as particles do not cross cell boundaries.
Furthermore, the load balance improves because particles with similar
velocities will have similar orbit lengths, thus improving the efficiency
of the algorithm at the warp level. This approach needs two copies
of the particle array (corresponding to timesteps $n$ and $n+1$),
and the computational complexity of the sorting scheme scales with
the number of particles. However, only a single sorting step is required
per implicit timestep, making the overall overhead manageable.

\subsection{Control-flow optimization}

The implicit particle mover uses control flows extensively to ensure
an accurate orbit integration with large timesteps. In this case,
performance may degrade on the GPU because warp execution results
in branches, which are executed sequentially. 

Branches are created, for instance, when a particle crosses a cell
boundary, or a particle orbit terminates, or memory accumulations
collide, etc. Some optimizations introduced in the preceding sections
already address branching and load balancing, e.g., by solving the
Crank-Nicholson equations directly (see Sec. \ref{sub:Algorithmic-element-2-CN}),
and by reducing memory collisions (see Sec. \ref{sub:Particle-current-deposition}).
Particle cell-crossings introduce branches because they happen randomly
in a given thread, thus forcing other threads in the same warp to
wait. The quantitative performance impact of particle cell-crossing
is shown in the next section. 

Branches also appear when particle orbits within a warp do not terminate
simultaneously. Their impact is ameliorated in our implementation
by using the vote function with \texttt{all} reduction mode \citep{nvprogramming}
as the exit condition for a given warp. The \texttt{vote.all} function
returns \texttt{true} when the exit condition is satisfied for all
the threads in a warp, and it can be used to break an infinite while-loop
(which does not create conditional branches). We have found that the
warp vote function is more effective for VP than for VA. This is most
likely due to the fact that increased atomic operations of the VA
approach hide load-imbalance latencies.

\section{Numerical experiments}

\label{sec:Num-Examples}

In this section, we first conduct several numerical experiments to
characterize the baseline and optimized particle orbit integration
algorithms described above. We measure their operational throughput
on the GPU, and demonstrate the significant performance and efficiency
gains of the proposed optimizations, resulting in a 50-70\% intrinsic
efficiency (vs. the application maximum operational throughput) for
VA and VP, respectively, with a corresponding 20-25\% overall efficiency
(vs. peak throughput, 1581 GOp/s). We also study their performance
sensitivity to various external parameters such as the number of threads
and the timestep size, and conclude that performance is generally
robust except for the timestep size (which directly affects the OI
of the algorithm). We compare the performance of the particle pusher
algorithm running on both a CPU (in single precision) and a GPU (in
single precision), and demonstrate significant speedups $(200-300)$.
Finally, we integrate the GPU particle mover with the full nonlinear
solver \citep{Chen20117018} on a CPU, and test accuracy and wall-clock
performance for a challenging ion acoustic wave simulation. We demonstrate
that a mixed-precision hybrid CPU-GPU implementation (with the GPU
running in single precision and the CPU in double precision) is sufficient
from an accuracy standpoint, and that it is able to deliver very large
wall clock speedups vs. the double-precision CPU-only implementation.
In what follows, code running on the GPU is in single-precision unless
otherwise specified.

\subsection{GPU performance of the implicit particle mover}

\begin{table}
\begin{centering}
\caption{\label{tab:avg-Op-Throughput}Breakdown of operations and clock cycles
(per thread per sub-step) of the implicit particle mover algorithm.
In the table, the left number counts operations, and the right one
counts clock cycles.}

\par\end{centering}

\centering{}%
\begin{tabular}{|>{\centering}p{0.8in}|>{\centering}m{1in}|>{\centering}m{1in}|>{\centering}m{0.8in}|>{\centering}m{0.8in}|>{\centering}m{0.9in}|}
\hline 
OP/CC & FP add,mul,

fma & AL add, mul, 

logic, cvt  & SF rsqrt, \_\_fdividef & division & total\tabularnewline
\hline 
baseline & 64.2 $/$ 48.0 & 31.8 $/$ 63.6 & 6.2 $/$ 49.6 & 2.1 $/$ 67.2 & 104 $/$ 228.4\tabularnewline
\hline 
optimized & 60.1 $/$ 43.5 & 26.6 $/$ 53.2 & 2.2 $/$ 17.6 & 0 $/$ 0 & 88.9 $/$ 114.3\tabularnewline
\hline 
\end{tabular}
\end{table}

The GPU performance of the algorithm is best understood by comparing
the theoretical operational throughput to that of a real execution.
To compute the theoretical operational throughput, we adopt operations
per second (Op/s), instead of FLOPS, as the figure of merit for performance
evaluations. This is because floating-point, integer, logic, and special-function
operations play important roles in our algorithm. When computing the
operational throughput, we need count all operations, especially the
slower ones. 

In the simulations, we set $\varepsilon_{a}=10^{-8}$ and $\varepsilon_{r}=0.02$
to be the absolute and relative tolerances of the orbit sub-time-step
estimation, respectively (see Sec. \ref{sub:Algorithmic-element-1-sub-time}).
The corresponding average number of cell-crossings is about 10\% of
the number of sub-steps. Table \ref{tab:avg-Op-Throughput} lists
the theoretical number of arithmetic operations and their respective
clock cycles for the baseline and optimized algorithms. We see that,
even though the total number of operations does not change much from
the baseline to the optimized version, there is a dramatic change
in the number of clock cycles. By examining the operational throughput
of each category, we see that the baseline algorithm spends most of
the time computing only a small number of special functions and divisions,
whereas the optimized algorithm significantly reduces the use of those
operations. As a result, the theoretical performance nearly doubles
after the optimization.

\begin{figure}
\includegraphics[width=\columnwidth]{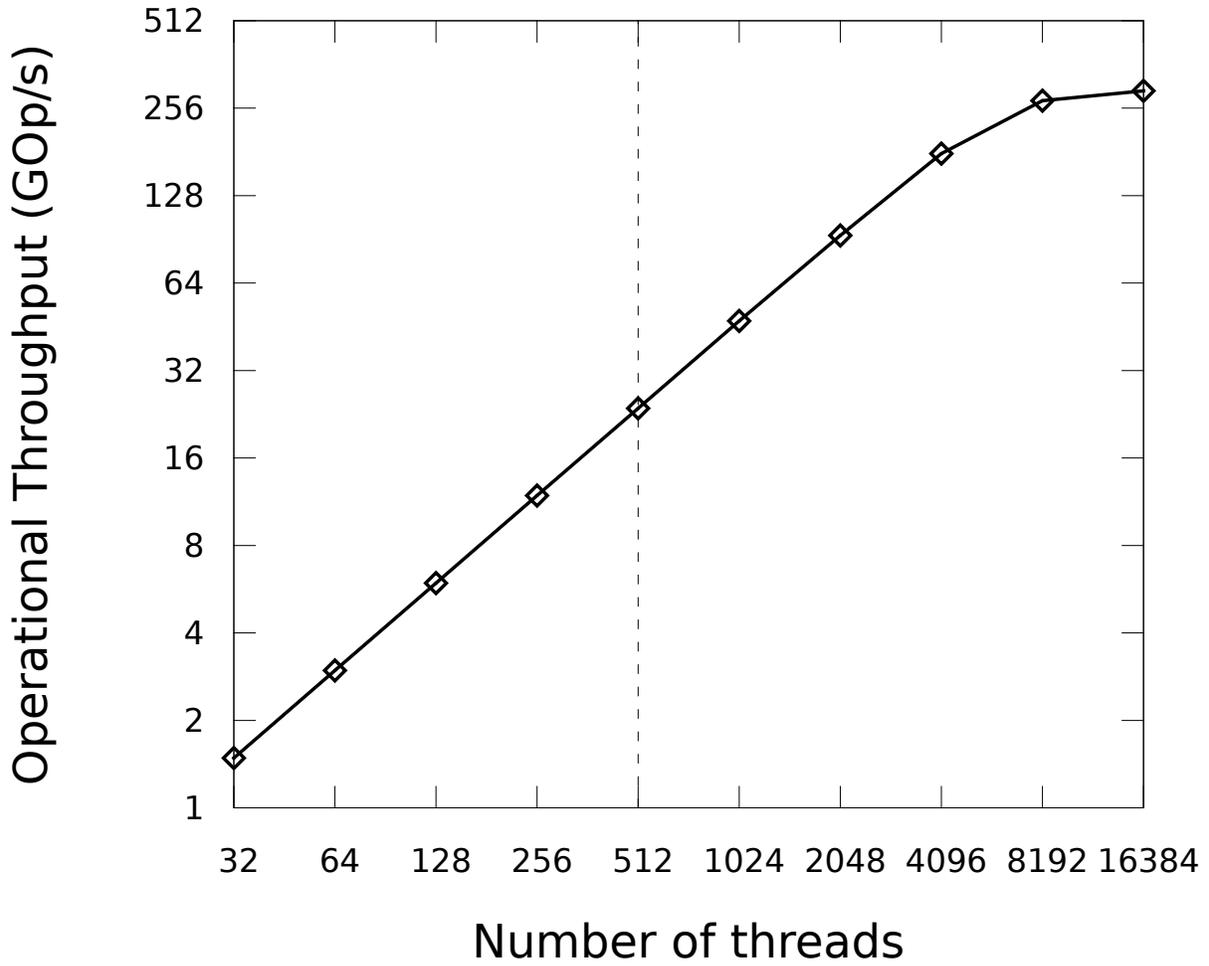}\caption{Scaling\label{fig:Scaling-of-ACC-CN} of implicit particle mover (optimized
for VA). The dotted line indicates the number of parallel CUDA cores
available on the GPU. The peak performance reaches 320 GOp/s. }
\end{figure}

Figure \ref{fig:Scaling-of-ACC-CN} shows the operational throughput
of the algorithms running on the GPU as a function of the number of
threads. We see that perfect linear scaling continues beyond the physical
number of CUDA cores, and performance only saturates when the number
of threads exceeds the number of cores by a factor of about 20. This
is consistent with Little's law\citep{little1961proof}, which predicts
that the number of threads needed for maximum performance is equal
to the number of cores multiplied by the instruction latency ($\sim18$
CC on Fermi GPUs\citep{nvbestpracticesguide}). Achieving the maximum
performance of Little's law is possible when 1) latencies, including
memory-level and instruction-level ones, are hidden by exploiting
extreme concurrency with many threads, and 2) the architecture allows
very fast switching between threads (GPUs can switch in one CC).

As implemented, the measured performance of the optimized algorithm
(for VA) is 320 GOp/s, which corresponds to a 50\% intrinsic efficiency
and a 20\% peak efficiency. This is to be compared with 130 GOp/s
of the baseline algorithm (8\% of peak). For the VP approach, we get
380 GOp/s (70\% intrinsic efficiency, 24\% peak efficiency). The performance
increase of VP vs. VA can be traced to the much larger number of accumulations
required by VA.

\begin{figure}
\includegraphics[width=\columnwidth]{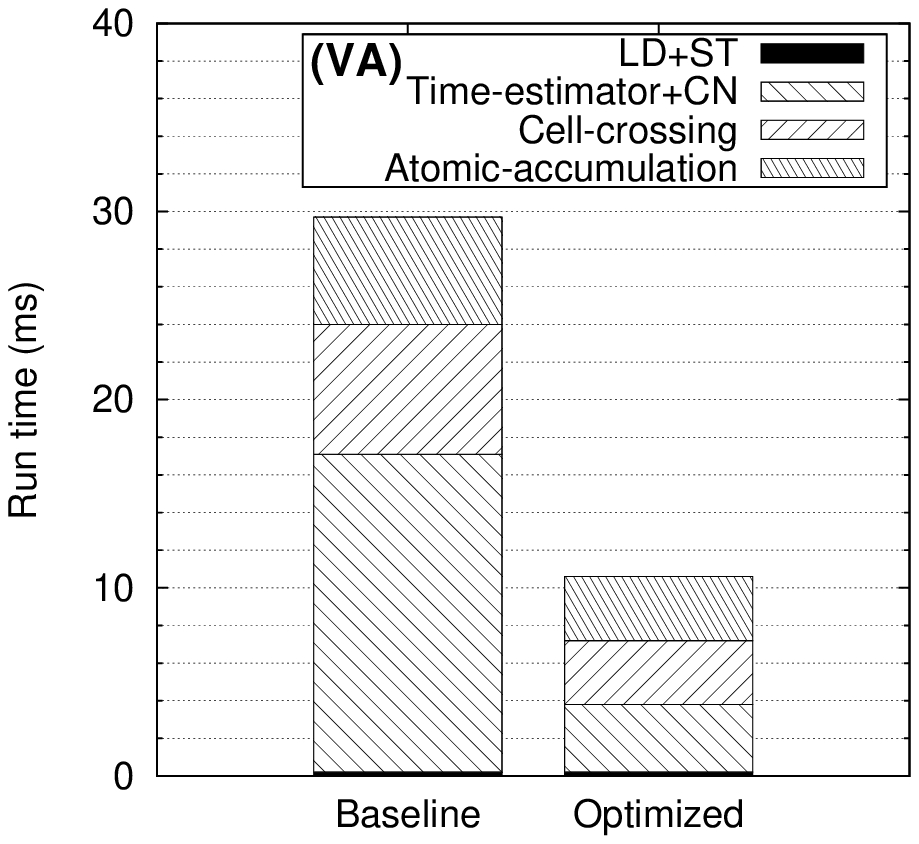}\caption{\label{fig:Run-time-breakdown}Run-time breakdown of the mover algorithm
for the VA approach. The load (LD) and store (ST) of 1,048,576 particle
quantities on the global memory take 0.2 ms. The VP algorithm features
similar timings, except for the atomic-accumulation step (which becomes
negligible).}
\end{figure}

Figure \ref{fig:Run-time-breakdown} shows the run-time breakdown
of the algorithm before and after the optimizations. The cost of global
memory operations is negligible compared to other operations, confirming
that the algorithm is compute-bounded. The most time-consuming parts
are the time estimator and the Crank-Nicolson mover. They achieve
a significant speedup, a result of both thread-level (through modifications
of the algorithm) and warp-level (through particle sorting and vote
function) optimizations. The optimizations in the cell-crossing algorithmic
element are also effective. 

\begin{figure}
\includegraphics[width=\columnwidth]{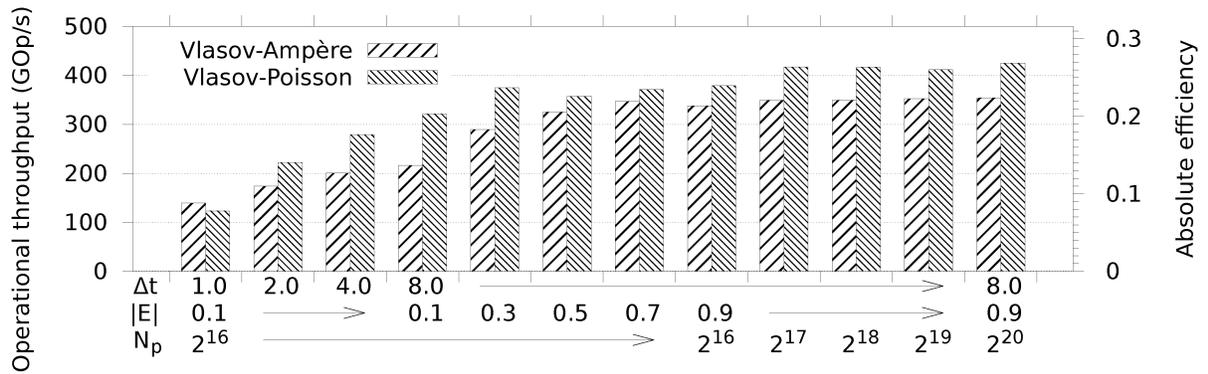}\caption{\label{fig:Performance-and-efficiency}Performance and efficiency
of the mover algorithm in single precision on the GPU under conditions
varying the timestep $(\Delta t)$, the field amplitude ($|E|$),
and the number of particles ($N_{p}$). For large timesteps, the performance
is insensitive to the field strength or the particle number.}
\end{figure}

Figure \ref{fig:Performance-and-efficiency} shows the sensitivity
of the performance and efficiency of the optimized particle mover
algorithm for both the VA and VP approaches with respect to timestep,
field strength, and number of particles. Clearly, while the performance
is insensitive to changes in the field strength and the particles
number, it is quite sensitive to changes in the timestep, improving
with increasing $\Delta t$. For a timestep of 0.1 (typical in explicit
PIC simulations), the algorithm is close to a memory-bounded regime,
and the performance is low. As the timestep increases from 1 to 10
or larger (typical in implicit PIC simulations \citep{Chen20117018}),
the algorithm becomes compute-bounded, and both the VA and the VP
recover good performance as well as efficiency.

\subsection{Performance comparison between the CPU and GPU implementations of
the particle mover algorithm}

For compute-bounded algorithms, a CPU-GPU performance comparison is
informative when the speedup is measured against the theoretical peak
performance speedup. For a single Intel Xeon CPU X5460 core at 3.16
GHz (used in this work), the peak theoretical performance for executing
2 operations on 4 single-precision variables in the SIMD (single instruction,
multiple data) style per clock cycle is: 
\[
2\,\mathrm{SIMD\, Op/}\mathrm{CC}\times4\mathrm{\, value/Op}\times\mathrm{clock\, rate}=25.2\,\mathrm{GOp/s}.
\]
Therefore, assuming the same efficiency on both architectures, the
nominal GPU-to-CPU speedup is around $60\,(\simeq1581/25.2)$. 

\begin{figure}
\includegraphics[width=\columnwidth]{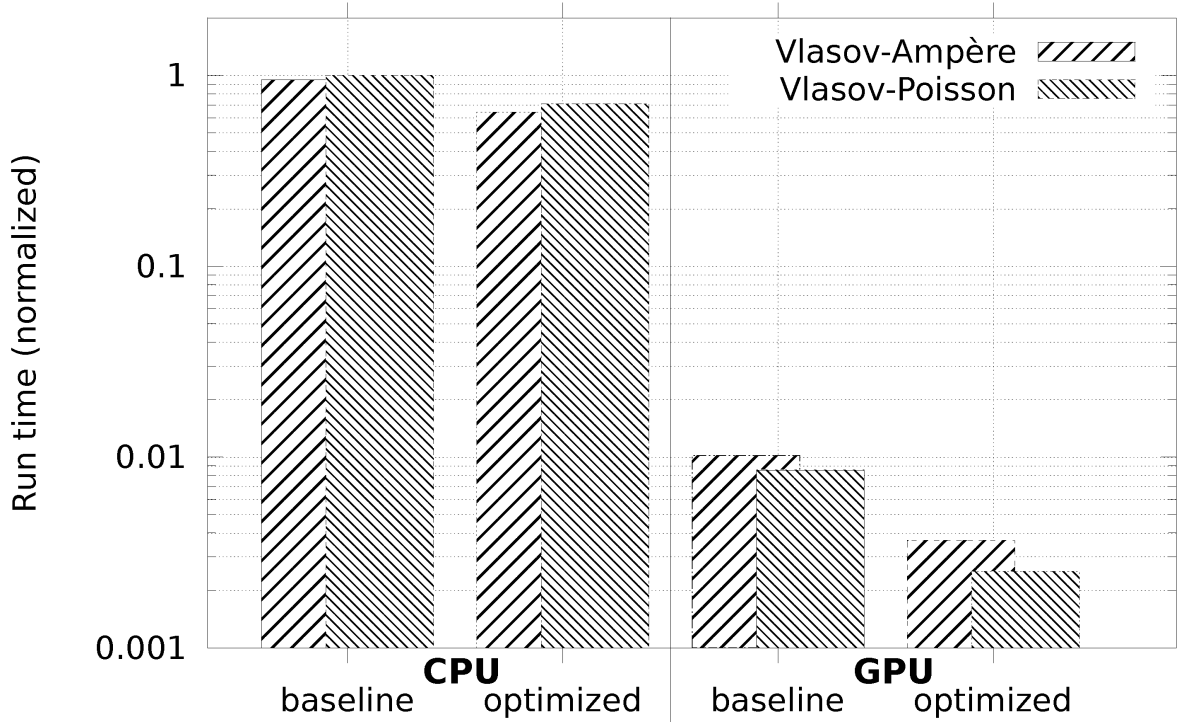}

\caption{\label{fig:CPU-and-GPU-runtime}CPU and GPU run time comparison for
the baseline and optimized mover algorithm of VA and VP approaches
(log scale). The single-precision GPU versions are two orders of magnitude
faster than the respective single-precision CPU versions. }
\end{figure}

Actual CPU vs. GPU timing comparisons are depicted in Fig. \ref{fig:CPU-and-GPU-runtime}.
We have programmed the CPU code with standard C/C++, without implementing
any explicit Intel SSE SIMD instructions, or multi-threading using
multi-cores. We have relied on the compiler (Intel C/C++ compiler
version 12.0.4) to optimize automatically. We first compare the performance
of the baseline algorithm between the CPU and GPU (this case is identical
to the one shown in Fig. \ref{fig:Run-time-breakdown}). The codes
are very similar on both processors, except for fast memory management,
which is explicit on the GPU (Sec.\eqref{sec:ACC-CN mover}). We find
that the absolute GPU efficiency ($\sim8\%$) is larger than that
of the CPU code ($\sim5\%$). The speedup scores 100. The increased
speedup (100 vs 60) is consistent with the increase in efficiency
(8\% vs 5\%). 

After the optimizations, the speedup reaches about 200 and 300 for
the VA and VP approaches, respectively, corresponding to a GPU efficiency
of 20-25\%. We note that some of the modifications introduced in Sec.\ref{sec:ACC-CN mover}
(such as the optimized sub-timestep estimator) have also been used
to improve the performance on the CPU.

\subsection{Performance of the hybrid, mixed-precision CPU-GPU fully implicit
PIC solver}

\begin{figure}
\includegraphics[width=\columnwidth]{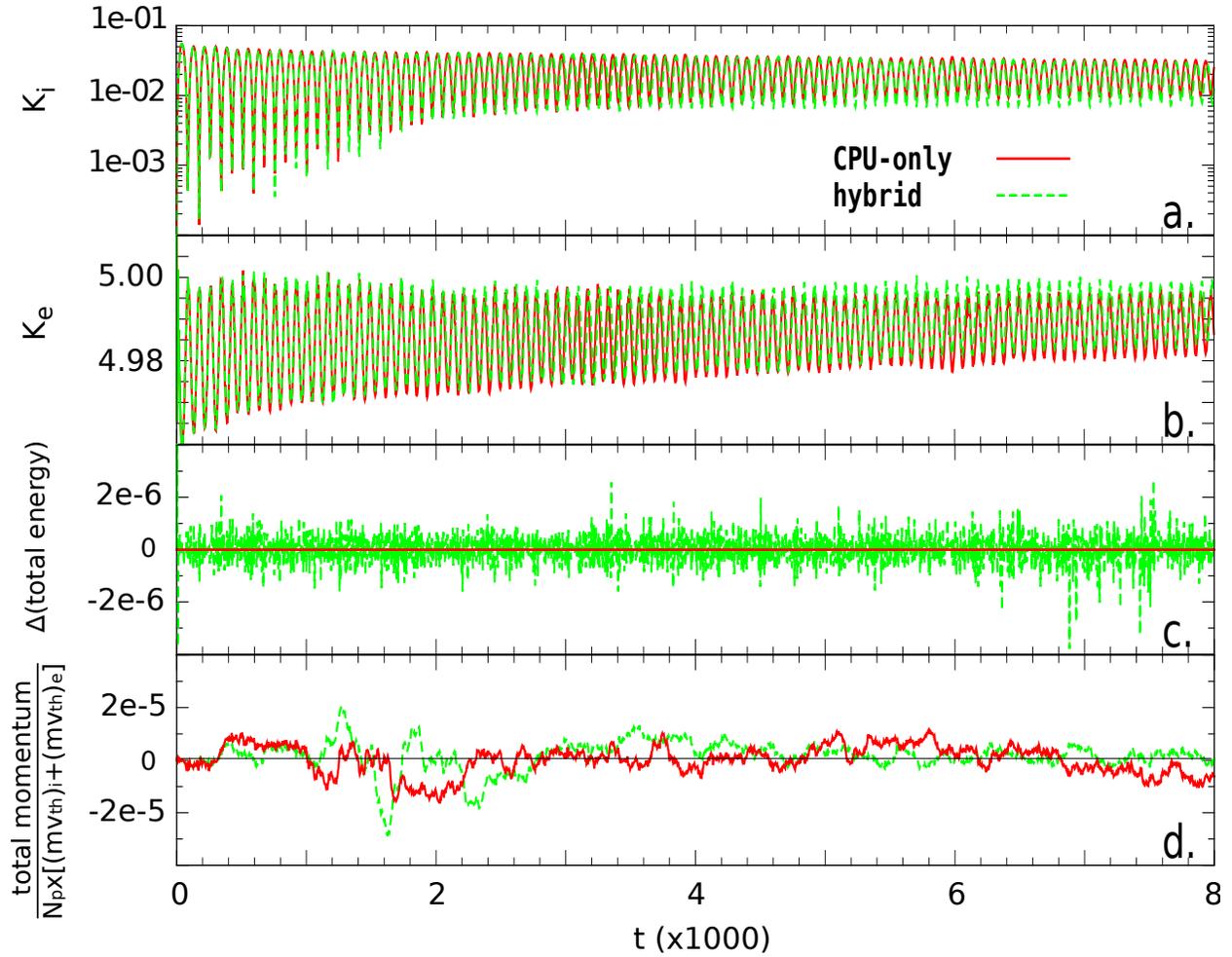}

\caption{\label{fig:Long-time-simulation-iaw}Long-timescale simulation of
the IAW problem, comparing simulations with the particle mover on
the CPU using double precision and on the GPU using single precision.
Panel a,b,c,d are the time history of ion kinetic energy, electron
kinetic energy, total energy variation every timestep, and normalized
total momentum, respectively. }
\end{figure}

We proceed to compare the performance of a hybrid, mixed-precision
CPU-GPU implementation of the fully implicit PIC algorithm vs. the
CPU-only serial implementation in Ref. \citep{Chen20117018} (but
incorporating applicable algorithmic optimizations developed in this
study). We follow this reference and use the ion acoustic wave (IAW)
case for our numerical tests. As set up, the IAW features large-amplitude
waves that can propagate in an unmagnetized, collisionless plasma
without significant damping. 

Figure \ref{fig:Long-time-simulation-iaw} shows VA simulation results
from both the hybrid implementation and the CPU-only one. We depict
the ion and electron kinetic energy, as well as conserved quantities
(local charge, total momentum, and total energy). The relative nonlinear
tolerance of the JFNK solver is $2\times10^{-4}$ for the hybrid CPU-GPU
version, and $10^{-10}$ for the CPU-only version. The nonlinear tolerance
is larger in the GPU version to prevent stalling of the nonlinear
iteration due to lack of numerical precision in the calculation of
the current density. As a result, the average number of Newton iterations
per timestep is about 5 for the single precision implementation, vs.
8 for the double-precision one. 

Figure \ref{fig:Long-time-simulation-iaw} demonstrates very good
agreement between the two implementations over a very long simulation
span (100 IAW periods or 8000 plasma periods). Conservation of energy
is enforced to the nonlinear tolerance level, i.e., $10^{-12}$ and
$10^{-6}$ for double- and single-precision simulations, respectively.
The total momentum is not conserved exactly in either implementation,
but it remains bounded and fluctuates at similar levels in both simulations.
The wall clock comparison of the two simulations shows a speedup of
over 130. The speedup reduces to 70 when the same nonlinear tolerance
($2\times10^{-4}$) is employed in both CPU-GPU and CPU-only simulations.
These speedup factors are consistent with Amdahl's law, as the particle
mover represents $\sim$98\% of the overall cost of the algorithm
in the CPU-only implementation. Larger speedups are possible for problem
setups where the particle cost represents a larger fraction. 

We have retrofitted the mixed-precision, hybrid CPU-GPU implementation
with a defect-correction algorithm \citep{defect-correction} that
delivers a true double precision solution. In this version of the
mixed-precision hybrid algorithm, the nonlinear residual in Newton's
method is evaluated in double precision, while the Jacobian-vector
product are evaluated in single precision. For the IAW problem, this
hybrid implementation requires only about a third of the GPU particle
mover calls to be evaluated in double precision. It is important to
note that double-precision computations in our GPU target architecture
are expensive, not only due to a factor of 8 (or 2 in some of the
newest GPUs) cost increase of double-precision operations, but also
because current GPU architectures do not yet feature double-precision
versions of the fast intrinsic functions or atomic operations. Nevertheless,
our defect-correction implementation of the mixed-precision hybrid
algorithm is still able to achieve a speedup of 40, which is 1.6 times
better than a full double-precision hybrid CPU-GPU computation.

\section{Discussion and Conclusions}

\label{sec:Conclusion}

This study has explored the hybrid, mixed-precision implementation
of a recently proposed fully implicit PIC algorithm \citep{Chen20117018}
on a heterogeneous (CPU-GPU) architecture. The implicit PIC algorithm
is ideally suited for a hybrid implementation owing to the concept
of particle enslavement, which segregates the particle orbit integration
in the nonlinear residual evaluation. Accordingly, the particle mover
can be farmed off to a GPU, while the rest of the nonlinear solver
machinery remains on the CPU. 

With the aid of the roofline model, we have demonstrated that the
implicit adaptive, charge-conserving particle mover algorithm of Ref.
\citep{Chen20117018} is compute-bounded, unlike memory-bounded explicit
particle movers \citep{decyk2010adaptable,kong2010particle,burau2010picongpu,madduri2011gyrokinetic}.
The optimized parallelization of the particle mover on the GPU significantly
boosts the performance compared to both the algorithm's original CPU
implementation and a straightforward GPU implementation. The optimized
particle mover exploits the powerful floating-point units and fast
special-function units on the GPU without loss of accuracy. Significant
acceleration has been achieved by eliminating the very-expensive IEEE
divisions and square-roots, and by minimizing the creation of divergent
branches. We have adopted a novel particle sorting strategy that sorts
particles according to both their positions and their velocities to
improve memory efficiency and load balancing. The operational throughput
of the optimized particle mover algorithm reaches 300-400 GOp/s (for
VP and VA, respectively), on the Nvidia GTX 580 GPU in single precision
and with typical (large) implicit timesteps, corresponding to 50-70\%
intrinsic efficiencies (vs. the maximum algorithmic throughput) and
20-25\% overall efficiencies (vs. peak throughput).

Moment accumulations from particles are often quoted as a main bottleneck
in many previous explicit particle mover algorithms \citep{stantchev2008fast,madduri2009memory,burau2010picongpu}.
In contrast, the implicit mover algorithm spends most of the time
pushing particles. This is true even for the VA approach, which requires
memory accumulations at every sub-timestep. We find that the atomic
accumulations in the VA approach take about 30\% run time of the mover
algorithm. If the VP approach is adopted instead, the accumulation
overhead becomes almost negligible. For the other parts of the mover
algorithm, we find that the time spent in particle cell-boundary-stopping
(needed for exact local charge conservation) is comparable to that
in the sub-timestep estimate and the Crank-Nicolson mover.

Significant speedup of the whole implicit PIC algorithm is achieved
by the hybrid, mixed-precision CPU-GPU implementation (using single
precision in the GPU and double precision in the CPU) vs. a CPU-only
one (using double-precision) on a challenging multiscale test problem,
the ion acoustic wave. Speedups about 100 are found, which are consistent
with Amdahl's law. Careful comparison of relevant quantities (local
charge, energy, momentum, and the electron/ion kinetic energy) shows
very good quantitative agreement. Particularly encouraging is the
fact that errors in momentum conservation seem unaffected by the mixed-precision
character of the hybrid implementation. Similarly, JFNK performance
seems unaffected by the use of single precision particle computations
on the GPU, as long as the nonlinear tolerance is adjusted accordingly.
Overall, the mixed-precision hybrid implementation is found to provide
a robust enough algorithm for this simulation. 

A mixed-precision CPU-GPU implementation that delivers a true double-precision
simulation capability has also been implemented using a defect-correction
approach. The speedup (vs. the CPU-only double precision serial simulation)
is about 40. This outperforms a full double precision CPU-GPU implementation,
which results in a speedup of 25. 

This study demonstrates at a proof-of-principle level that hybrid
CPU-GPU implementations hold much promise for future investigation
in the context of fully implicit PIC algorithms. Future work should
focus on extending the approach to multiple dimensions and larger
domains. With current limitations in GPU shared memory (a Fermi GPU
currently has at most 48KB of shared memory per SM), larger problem
sizes will require the use of domain decomposition among GPU nodes.
Given that the particle orbit integrator in our implicit PIC algorithm
may sample a relatively large fraction of the domain, special attention
will need to be paid to devise strategies to minimize communication
overhead. Solutions to these issues will be explored in future work.

\section*{Acknowledgments}

The authors would like to thank Dr. David L. Green for his help with
the Fortran-C/C++ mixed language implementation. This work has been
funded by the Oak Ridge National Laboratory (ORNL) Directed Research
and Development program (LDRD). ORNL is operated by UT-Battelle for
the US Department of Energy under contract DE-AC05-00OR22725.

\newpage{}

\bibliographystyle{elsarticle-num}
\bibliography{gpu}

\end{document}